\documentclass[reqno,centertags]{amsart}
\usepackage{amsmath,amsthm,amscd,amssymb}
\usepackage{latexsym}

\newcommand{\bbR}{{\mathbb{R}}}
\newcommand{\bbZ}{{\mathbb{Z}}}

\newcommand{\lb}{\label}
\newcommand{\f}{\frac}
\newcommand{\wti}{\widetilde  }
\newcommand{\tr}{\text{\rm{Tr}}}
\newcommand{\dist}{\text{\rm{dist}}}
\newcommand{\crit}{\text{\rm{crit}}}

\newcommand{\ess}{\text{\rm{ess}}}

\newcommand{\supp}{\text{\rm{supp}}}
\newcommand{\bi}{\bibitem}
\newcommand{\beq}{\begin{equation}}
\newcommand{\eeq}{\end{equation}}
\newcommand{\ba}{\begin{align}}
\newcommand{\ea}{\end{align}}
\newcommand{\veps}{\varepsilon}

\newcounter{smalllist}
\newenvironment{SL}{\begin{list}{{\rm\roman{smalllist})}}{%
\setlength{\topsep}{0mm}\setlength{\parsep}{0mm}\setlength{\itemsep}{0mm}%
\setlength{\labelwidth}{2em}\setlength{\leftmargin}{2em}\usecounter{smalllist}%
}}{\end{list}}

\DeclareMathOperator{\Real}{Re}

\allowdisplaybreaks
\numberwithin{equation}{section}

\newtheorem{theorem}{Theorem}[section]
\newtheorem*{t1}{Theorem 1}
\newtheorem*{t2}{Theorem 2}
\newtheorem*{t3}{Theorem 3}
\newtheorem*{c4}{Corollary 4}
\newtheorem*{t5}{Theorem 5}
\newtheorem*{p2.1}{Proposition 2.1}
\newtheorem{proposition}[theorem]{Proposition}
\newtheorem{lemma}[theorem]{Lemma}
\newtheorem{corollary}[theorem]{Corollary}
\theoremstyle{definition}

\newtheorem{example}[theorem]{Example}

\theoremstyle{remark}

\newcommand{\abs}[1]{\lvert#1\rvert}

\begin{document}
\title[Variational Estimates for Discrete Schr\smash{\"o}dinger Operators]{Variational Estimates for \\
Discrete Schr\"odinger Operators with \\ Potentials of Indefinite Sign}
\author[D. Damanik, D. Hundertmark, R. Killip, and B. Simon]{D. Damanik$^{1,3}$, 
D. Hundertmark$^2$, R. Killip$^1$, and B. Simon$^{1,4}$}

\thanks{$^1$ Mathematics 253-37, California Institute of Technology, Pasadena, CA 91125. 
E-mail: damanik@its.caltech.edu; killip@its.caltech.edu; 
bsimon@caltech.edu}
\thanks{$^2$ Institut Mittag-Leffler, Aurav\"agen 17,  
S-182 60 Djursholm, Sweden.
On leave from
Department of Mathematics, 
University of Illinois at Urbana-Champaign, 
1409 W.~Green Street, 
Urbana, IL 61801-2975.}
\thanks{$^3$ Supported in part by NSF grant DMS-0227289}
\thanks{$^4$ Supported in part by NSF grant DMS-0140592}

\date{November 5, 2002}

\begin{abstract} Let $H$ be a one-dimensional discrete Schr\"odinger operator. We prove that if 
$\sigma_{\ess} (H)\subset [-2,2]$, then $H-H_0$ is compact and $\sigma_{\ess}(H)=[-2,2]$. 
We also prove that if $H_0 + \f14 V^2$ has at least one bound state, then the same is true for $H_0 +V$.
Further, if $H_0 + \f14 V^2$ has infinitely many bound states, then so does $H_0 +V$.
Consequences include the fact that for decaying potential $V$ with $\liminf_{|n|\to\infty} |nV(n)| > 1$,
$H_0 +V$ has infinitely many bound states; the signs of $V$ are irrelevant. Higher-dimensional analogues 
are also discussed.
\end{abstract}

\maketitle

\section{Introduction} \lb{s1}

Let $H$ be a Schr\"odinger operator on $\ell^2 (\bbZ)$,
\begin{equation}
(H u)(n) = u(n+1) + u(n-1) + V(n) u(n) \lb{1.3} 
\end{equation}
with bounded potential $V:\bbZ\to\bbR$.  The free Schr\"odinger operator, $H_0$, corresponds
to the case $V=0$.
One of our main results in this paper is 

\begin{t1} If $\sigma_{\ess}(H)\subset [-2,2]$, then $V(n)\to 0$ as $|n|\to\infty$,
that is, $H-H_0$ is compact.
\end{t1}

{\it Remark.} By Weyl's Theorem, we have the immediate corollary that $\sigma_{\ess}(H)=[-2,2]$
if and only if $V(n)\to 0$.

\medskip

Our motivation for this result came from two sources:

\begin{t2}[Killip-Simon \cite{KS}] If $\sigma (H)\subset [-2,2]$, then $V=0$. 
\end{t2}

\begin{t3}[Rakhmanov \cite{Rak}; see also Denisov \cite{Den}, Nevai \cite{Nev}, and references 
therein] Let $J$ be a general half-line Jacobi matrix on $\ell^2 (\bbZ^+)$,
\begin{equation} \lb{1.3a}
(Ju)(n)=a_n u(n+1) + b_n u(n) + a_{n-1} u(n-1)
\end{equation}
where $a_n>0$ and $\bbZ_+=\{1,2,\dots\}$. Suppose that $[-2,2]$ is the essential support of the 
a.c.~part of the spectral measure and also the essential spectrum. Then $\lim_{n\to\infty} \abs{a_n -1} 
+ \abs{b_n} =0$, that is, $J$ is a compact perturbation of $J_0$, the Jacobi matrix with $a_n \equiv 1$, 
$b_n \equiv 0$.
\end{t3}

While Theorem~3 motivated our thoughts, it is not closely related to the result. Not only are
the methods different, but it holds for any a priori $a_n$; whereas our results require some 
a priori estimates like $a_n\to 1$ as $\abs{n}\to\infty$. For example, if $a_n\equiv \f12$ and 
$b_n$ takes values $+1$ and $-1$ over longer and longer intervals, it is not hard to see that 
$\sigma (J) =[-2,2]$, but clearly, $J-J_0$ is not compact. Thus Theorem~1, unlike Theorem~3, is 
essentially restricted to discrete Schr\"odinger operators.

For continuum Schr\"odinger operators, consideration of sparse positive nondecaying potentials shows that
$\sigma(H)=[0,\infty)$ is possible even when $(H+1)^{-1} - (H_0 +1)^{-1}$ is not compact.
The reason is that our proof depends essentially---as does 
Theorem~2---on the fact that $\sigma (H)$ has two sides in the discrete case. 

Theorem~1 has an interesting corollary:

\begin{c4} Let $H$ be an arbitrary one-dimensional discrete Schr\"odinger operator. Then $\sup 
\sigma_{\ess}(H) - \inf\sigma_{\ess}(H)\geq 4$ with equality if and only if $V(n)\to V_\infty$ a 
constant as $\abs{n}\to\infty$. 
\end{c4}

\begin{proof} Let $a_+=\sup\sigma_{\ess}(H)$, $a_- =\inf\sigma_{\ess}(H)$. If $a_+ - a_- 
\leq 4$, then $H-\f12 (a_+ + a_-)$ is a Schr\"odinger operator with essential spectrum in $[-2,2]$.  
So Theorem~1 implies the original $V(n)\to\f12 (a_+ +a_-)$. Hence, $a_+ -a_- =4$ and $\sigma_{\ess}
=[a_-, a_+]$. 
\end{proof}

{\it Remarks.} (a) A similar argument combined with Theorem~2 implies that if $\sup\sigma (H)-\inf\sigma 
(H)\leq 4$, then $V$ is a constant. 

\smallskip
(b) If $V(n)=(-1)^n \lambda$ and $\lambda$ is large, standard Floquet theorem arguments show 
that $\sigma (H)$ has two bands centered about $\pm (\lambda +O(\f1{\lambda}))$ and of width 
$O(\f1{\lambda})$. Thus, while the size of the convex hull of $\sigma (H)$ is of size at 
least $4$, the size of $\sigma (H)$ can be arbitrarily small. Indeed, by results of 
Deift-Simon \cite{DS}, if $H$ has purely a.c.~spectrum, (e.g., $V$ periodic), the total size 
of $\sigma (H)$ is at most $4$.

\medskip
While Theorem~1 is our main motivating result, the ideas behind it yield many other results about 
the absence of eigenvalues and about the finiteness or infinitude of their number for Schr\"odinger operators
not only on the line, but also on the half-line or in higher dimensions. Included in our results are 
\begin{SL} 
\item[(i)] Theorem~1 holds in two dimensions and is false in three or more dimensions (see 
Theorems~\ref{T4.1} and \ref{T4.2}). This is connected to the fact that Schr\"odinger operators in 
one and two dimensions always have a bound state for nontrivial attractive potentials 
(see \cite[pp.~156--157]{LL} and \cite{Klaus,Simon76}), whereas in three and more dimensions, small 
attractive potentials need not have bound states by the Cwikel-Lieb-Rozenblum bound \cite{Cwi,Lieb,Roz}. 
\item[(ii)] For a half-line discrete Schr\"odinger operator, $H$, if $\sigma (H)=[-2,2]$ (i.e., no 
bound states), then (see Theorem~\ref{T5.2}) 
\begin{equation} \lb{1.4}
\abs{V(n)} \leq 2n^{-1/2}
\end{equation}
On the other hand (see Theorem~\ref{T5.2}), there are examples, $V_k(n)$, with no bound states and 
$\lim_k \sup_n n^{1/2} \abs{V_k(n)}=1$. This shows that the power $\f12$ in \eqref{1.4} cannot be 
made larger.  It also shows that the constant, $2$, cannot be made smaller than $1$.
(The optimal constant is $\sqrt{2}$. This is proved in \cite{DK}.)
\item[(iii)] The examples in (ii) are necessarily sparse in that if $\abs{V(n)}\geq Cn^{-\alpha}$ 
and $H$ has only finitely many bound states, then $\alpha \geq 1$. Indeed, we will prove (see 
Theorem~\ref{T5.6}) that if $\alpha =1$ and $C>1$ or $\alpha <1$ and $C>0$, then $H$ has an 
infinity of bound states. This will follow from the very general theorem: 
\end{SL} 

\begin{t5} Let $V(n)\to 0$. If $H_0 + \f14 V^2$ has at least one 
{\rm{(}}resp., infinitely many{\rm{)}} eigenvalues outside $[-2,2]$, then $H_0 +V$ has at least 
one {\rm{(}}resp., infinitely many{\rm{)}} eigenvalues outside $[-2,2]$.
\end{t5} 

\begin{SL} 
\item[] Theorem~\ref{T3.1} extends this result to all dimensions.

\item[(iv)] If $\abs{V(n)}\geq Cn^{-\alpha}$ and $\alpha <1$, we will prove suitable eigenvalue 
moments diverge. 
\end{SL} 

\smallskip
The starting point of the present paper is the discussion at the end of Section~10 of \cite{KS} that 
it should be possible to prove Theorem~2 variationally with suitable second-order perturbation trial 
functions. Second-order eigenvalue perturbation theory has a change of the first-order eigenfunction 
by a term proportional to $V$\!. Thus, our variational trial function will have two pieces: $\varphi$ 
and an extra piece, proportional to $V\varphi$. 

The second key idea is to make use of the fact that the spectrum of $H_0$ has two sides, and we can 
use a pair of trial functions: one to get an eigenvalue below $-2$ and one to get an eigenvalue above 
$+2$. By combining them, we will have various cancellations that involve terms whose sign is uncertain. 
Explicitly, given a pair of trial vectors $\varphi_+$ and $\varphi_-$, we define
\begin{equation} \lb{1.5}
\Delta (\varphi_+, \varphi_-; V) = \langle \varphi_+, (H-2)\varphi_+\rangle + \langle \varphi_-, 
(-H-2)\varphi_-\rangle
\end{equation}
where $H$ is given by \eqref{1.3}. If $\Delta >0$, either $\langle \varphi_+, (H-2)\varphi_+\rangle >0$ 
or $\langle \varphi_-, (H+2)\varphi_-\rangle <0$, that is, there is either an eigenvalue above $2$ or 
below $-2$!

In choosing $\varphi_-$ relative to $\varphi_+$, it will help to use the unitary operator $U$ on $\ell^2 
(\bbZ)$ given by 
\begin{equation} \lb{1.6}
(U\varphi)(n) = (-1)^n \varphi(n)
\end{equation}
so that 
\begin{equation} \lb{1.7}
UH_0 U^{-1} = -H_0 \qquad UVU^{-1} =V
\end{equation}
The key calculation in Section~\ref{s2} will be that 
\begin{equation} \lb{1.8}
\Delta (\varphi + \tfrac14\, V\varphi, U(\varphi -\tfrac14\, V\varphi)) \geq 2 
\langle \varphi, [H_0 + \tfrac14\, V^2 -2]\varphi\rangle
\end{equation}

For example, this immediately implies the ``at least one bound state" part of Theorem~5. If $H_0 + \f14 
V^2$ has a bound state, $\varphi$, we must have $\langle \varphi, (H_0 + \f14 V^2)\varphi\rangle > 2 
\langle \varphi, \varphi\rangle$, so $\Delta >0$. 

The current paper complements \cite{DHS}.
That paper provided upper bounds on the distance from
$[-2,2]$ of eigenvalues of discrete Schr\"odinger operators with oscillatory potentials. This paper 
provides lower bounds. In particular, there it was shown the Jacobi matrix with $a_n\equiv 1$, $b_n =
\f{\beta(-1)^n}{n}$ has finitely many eigenvalues if $\abs\beta\leq\f12$. Here, we prove infinitely many 
(see Theorem~\ref{T5.7}) if $\abs\beta >1$.  We also show, by ad hoc methods, that there are no eigenvalues
for $\abs\beta\leq1$ (see Proposition~\ref{P:AltEx}).

In Section~\ref{s2}, we prove variational estimates, including \eqref{1.8}. In Section~\ref{s3}, we 
prove Theorem~5. In Section~\ref{s4}, we prove Theorem~1 and provide a new proof of Theorem~2. 
Sections~\ref{s2}--\ref{s4} also discuss higher dimensions. In Section~\ref{s5}, we study the 
one-dimensional situation more closely. 

\smallskip
We thank Andrej Zlato\v{s} for useful discussions. 

\bigskip
\section{Variational Estimates} \lb{s2}

On $\ell^2 (\bbZ^\nu)$, define $H_0$ by 
\begin{equation} \lb{2.1}
(H_0u)(n) = \sum_{\abs{j}=1} u(n+j)
\end{equation}
so
\begin{equation} \lb{2.2}
-2\nu \leq H_0 \leq 2\nu 
\end{equation}
For $V$\!, a bounded function on $\bbZ^\nu$, let 
\begin{equation} \lb{2.3}
H=H_0+V
\end{equation}
We are interested in the spectrum of $H$ outside $[-2\nu, 2\nu]=\sigma(H_0)$. 

If we define $U$ on $\ell^2 (\bbZ^\nu)$ by 
\begin{equation} \lb{2.4}
(U\varphi)(n) =(-1)^{\abs{n}}\varphi(n)
\end{equation}
where $\abs{n}=\abs{n_1} + \cdots + \abs{n_\nu}$, then
\begin{equation} \lb{2.5}
UH_0 U^{-1} =-H_0 \qquad UVU^{-1} =V
\end{equation}

We define, for $\varphi_+, \varphi_-\in \ell^2 (\bbZ^\nu)$, 
\begin{equation} \lb{2.6}
\Delta (\varphi_+, \varphi_-; V) = \langle \varphi_+, (H-2\nu) \varphi_+\rangle + 
\langle \varphi_-, (-H-2\nu)\varphi_-\rangle 
\end{equation}
$\Delta >0$ implies that $H$ has spectrum outside $[-2\nu, 2\nu]$ and, as we will see, $\Delta 
(\varphi_+^{(n)}, \varphi_-^{(n)};V)>0$ for suitable $\varphi_\pm^{(n)}$ implies the spectral 
projection $\chi_{\bbR\backslash [-2\nu, 2\nu]}(H)$ has infinite dimension.

Note first that 

\begin{proposition} \lb{P2.1} If $f,g\in \ell^2 (\bbZ^\nu)$, then 
\begin{equation} \lb{2.7}
\Delta (f+g, U(f-g); V) \geq 2\langle f, (H_0-2\nu)f\rangle - 8\nu \|g\|^2 + 4 \Real \langle f,
Vg\rangle
\end{equation}
\end{proposition} 

\begin{proof} By \eqref{2.5},  
\begin{align*}
\Delta (f+g, U(f-g);V) &= \langle (f+g), (H_0 -2\nu +V) (f+g)\rangle \\
&\qquad + \langle (f-g), (H_0-2\nu -V) (f-g)\rangle \\
&= 2\langle f, (H_0 -2\nu)f\rangle + 2 \langle g, (H_0 -2\nu)g\rangle + 4\Real \langle f, Vg\rangle 
\end{align*}
By \eqref{2.2}, $H_0 \geq -2\nu$, so 
\[
\langle g, (H_0 -2\nu)g\rangle \geq -4\nu \|g\|^2
\]
This yields \eqref{2.7}. 
\end{proof} 

One obvious choice is to take $f=\varphi$, $g=\gamma V\varphi$. The $V$-terms on the right side of 
\eqref{2.7} are then 
\begin{equation} \lb{2.8}
\|V\varphi\|^2 (-8\nu \gamma^2 + 4\gamma)
\end{equation}
which is maximized at $\gamma = \f{1}{4\nu}$, where $-8\nu\gamma^2 + 4\gamma =\f1{2\nu}$. Thus we have 
a generalization of \eqref{1.8}. 

\begin{theorem} \lb{T2.2} For any $\varphi\in \ell^2 (\bbZ^\nu)$,
\begin{equation} \lb{2.9}
\Delta\big(( 1+\tfrac{1}{4\nu} \, V)\varphi, U(1-\tfrac{1}{4\nu}\, V)\varphi; V\big) \geq 2 \langle \varphi, 
(H_0 -2\nu + \tfrac{1}{4\nu}\, V^2)\varphi\rangle
\end{equation}
\end{theorem} 

In some applications, we will want to be able to estimate $\|f\pm g\|$ in terms of $f$, and so want 
to cut off $Vg$. We have 

\begin{theorem} For any $F\in\ell^\infty$ with $0\leq F\leq 1$, we have
\begin{equation} \lb{2.10x}
\Delta\big(\varphi (1+ (4\nu)^{-1} FV), U\varphi (1- (4\nu)^{-1} FV); V\big) \geq 2 \langle \varphi, 
(H_0 -2\nu + (4\nu)^{-1} FV^2)\varphi\rangle
\end{equation}
\end{theorem}  

\begin{proof} By taking $g=\gamma FV\varphi$, $f=\varphi$, the $V$-terms in \eqref{2.7} are 
\begin{equation} \lb{2.10}
-8\nu \gamma^2 \|FV\varphi\|^2 + 4\gamma \langle V\varphi, FV\varphi_-\rangle
\end{equation}
in place of \eqref{2.8}. Since $0\leq F\leq 1$, we have $-F^2 \geq -F$, so 
\[
-\|FV\varphi\|^2 \geq -\langle V\varphi, FV\varphi\rangle
\]
and \eqref{2.10x} results. 
\end{proof} 

The properties of $H_0$ needed above are only \eqref{2.2} and \eqref{2.5}. If $J$ is a Jacobi matrix 
\eqref{1.3a} and $J_1$ is the Jacobi matrix with the same values of $a_n$ but with $b_n=0$, then 
$UJ_1U^{-1}=-J_1$. \eqref{2.2} is replaced by 
\begin{equation}\lb{2.12}
J_1 \geq -\alpha
\end{equation}
where
\begin{equation}\lb{2.13}
\alpha =\max_n \,(a_n + a_{n+1})
\end{equation}

One has 
\begin{theorem}\lb{T2.4} For any $\varphi\in\ell^2 (\bbZ_+)$, with $\varphi_\pm = (1\pm\gamma V) 
\varphi$ {\rm{(}}where $\gamma=(2+\alpha)^{-1}${\rm{)}}, we have
\begin{equation}\lb{2.14}
\langle\varphi_+, (J-2)\varphi_+\rangle + \langle U\varphi_-, (-2-J)U\varphi_-\rangle \geq 
2\langle\varphi, (J_1 -2 + \gamma b^2)\varphi\rangle
\end{equation}
\end{theorem}

\bigskip
\section{A $V^2$ Comparison Theorem} \lb{s3}

Our goal in this section is to prove the following extension of Theorem~5: 

\begin{theorem}\lb{T3.1} Let $V$ be defined on $\bbZ^\nu$. Let $V(n)\to 0$ as $\abs{n}\to\infty$. 
If $H_0 + (4\nu)^{-1} V^2$ has at least one eigenvalue {\rm{(}}resp., infinitely many{\rm{)}} outside 
$[-2\nu,2\nu]$, then so does $H_0 +V$\!.
\end{theorem} 

The key to this will be Theorem~\ref{T2.2}, but we will also need 

\begin{lemma} \lb{L3.2} Let $W\geq 0$ on $\bbZ^\nu$ with $W(n)\to 0$ as $\abs{n}\to\infty$. 
If $H_0 +W$ has infinitely many eigenvalues in $(2\nu, \infty)$, then we can find $\{\varphi_n
\}_{n=1}^\infty$ with $\langle \varphi_n, (H_0 +W) \varphi_n\rangle >2\nu \|\varphi_n\|^2$, so that 
each $\varphi_n$ has finite support and 
\begin{equation} \lb{3.1}
\dist \big(\supp(\varphi_n), \supp(\varphi_m)\big)\geq 2
\end{equation} 
for all $n\neq m$. 
\end{lemma} 

\begin{proof} Let $\Lambda_k =\{n\in\bbZ^\nu \mid \max_{i=1,\dots,\nu} \abs{n_i} \leq k\}$. We first 
claim that for every $k$, there exists $\psi$ with $\psi = 0$ on $\Lambda_k$ so that $\langle \psi, 
(H_0 +W) \psi \rangle >2\nu \|\psi\|^2$. For let $\widetilde H_0$ be $H_0$ with Dirichlet boundary 
conditions on $\partial \Lambda_k$, that is, dropping off-diagonal terms $H_{0,ij}$ with $i\in\Lambda_k$, 
$j\notin \Lambda_k$ or vice-versa. $\widetilde H_0 -H_0$ is finite rank, so $\widetilde H_0 +W$ has 
infinitely many eigenvalues in $(2\nu, \infty)$. But $\wti H_0+W$ is a direct sum of an operator on 
$\ell^2 (\Lambda_k)$ and one on $\ell^2 (\bbZ^\nu \backslash\Lambda_k)$. Since $\dim \ell^2 (\Lambda_k)
<\infty$, we can find $\psi\in \ell^2 (\bbZ^\nu \backslash \Lambda_k)$ so $\langle \psi, (H_0 +W)\psi
\rangle = \langle \psi, (\wti H_0 +W)\psi\rangle >2\nu \|\psi\|^2$. 

Now pick $\varphi_n$ inductively as follows. After picking $\{\varphi_n\}_{n=1}^N$, we have each 
$\varphi_n$ has finite support, so there is a $\Lambda_k$ with each $\varphi_n =0$ on $\bbZ^\nu 
\backslash\Lambda_k$, $n=1, \dots, N$. By the initial argument, pick $\psi_{N+1}$ vanishing on 
$\Lambda_{k+1}$ so that $\langle \psi_{N+1}, (H_0 +W) \psi_{N+1}\rangle >2\nu \langle \psi_{N+1}, 
\psi_{N+1}\rangle$ and $\psi_{N+1}=0$ on $\Lambda_{k+1}$. Let $\psi_{N+1}^{(m)}$ be finitely supported 
approximations to $\psi_{N+1}$ which vanish on $\Lambda_{k+1}$. By continuity, for some $m$, 
$\langle \psi_{N+1}^{(m)}, (H_0 +W) \psi_{N+1}^{(m)}\rangle >2\nu \langle \psi_{N+1}^{(m)}, 
\psi_{N+1}^{(m)}\rangle$. Pick $\varphi_{N+1} = \psi_{N+1}^{(m)}$. 
\end{proof}

\begin{proof}[Proof of Theorem~\ref{T3.1}] If $H_0 + (4\nu)^{-1} V^2$ has at least one eigenvalue 
outside $[-2\nu, 2\nu]$, there exists $\varphi$ with $\langle \varphi, (H_0 + \f1{4\nu}V^2 -2\nu)
\varphi\rangle >0$. By \eqref{2.9}, $H_0+V$ has some eigenvalue outside $[-2\nu, 2\nu]$. 

If $H_0 + (4\nu)^{-1} V^2$ has infinitely many eigenvalues, by Lemma~\ref{L3.2}, there exist $\varphi_n$ 
obeying \eqref{3.1} so that $\langle \varphi_n, (H_0 +\f14 V^2) \varphi_n\rangle >2\nu \|\varphi_n\|^2$. 
By \eqref{2.9}, 
we can find $\psi_n$  with either $\langle \psi_n, (H_0 + V)\psi_n\rangle >2\nu \|\psi_n\|^2$ or 
$\langle \psi_n, (H_0 +V)\psi_n\rangle <-2\nu \|\psi_n\|^2$ and $\supp (\psi_n)\subset\supp(\varphi_n)$. 
By \eqref{3.1}, we have
\[
\langle \psi_n, \psi_m\rangle =0 \quad \text{and} \quad \langle \psi_n, (H_0 +V)\psi_m\rangle =0 
\quad \text{for } n\neq m
\]
Thus, by the min-max principle, $H_0 +V$ has an infinity of eigenvalues in either $(2\nu, \infty)$ or 
$(-\infty, -2\nu)$. 
\end{proof} 

Using Theorem~\ref{T2.4} in place of Theorem~\ref{T2.2}, we get

\begin{theorem}\lb{T3.2} Let $J(\{a_n\}, \{b_n\})$ be the Jacobi matrix \eqref{1.3a}. Suppose $a_n\to 
1$ and $b_n\to 0$ so $\sigma_{\ess}(J) =[-2,2]$. Let $\alpha$ be given by \eqref{2.13} and $\gamma = 
(2+\alpha)^{-1}$. If $J(\{a_n\}, \{\gamma b_n^2\})$ has at least one eigenvalue {\rm{(}}resp., infinitely 
many{\rm{)}} in $(2,\infty)$, then $J(\{a_n\}, \{b_n\})$ has at least one eigenvalue 
{\rm{(}}resp., infinitely many{\rm{)}} in $(-\infty, -2) \cup (2,\infty)$.
\end{theorem} 

{\it Remark.} In particular, if $J(\{a_n\}, \{b_n=0\})$ has an infinity of eigenvalues, they cannot 
be destroyed by a crazy choice of $\{b_n\}$. 

\bigskip
\section{Essential Spectra and Compactness \\
in Dimension 1 and 2} \lb{s4}

Our goal in this section is to prove 

\begin{theorem}\lb{T4.1} Let $\nu=1$ or $2$. If $\sigma_{\ess}(H_0+V)\subset [-2\nu, 2\nu]$, then 
$V(n)\to 0$ as $\abs{n}\to\infty$. 
\end{theorem} 

\begin{theorem}\lb{T4.2} If $\nu \geq 3$, there exist potentials $V$ in $\ell^\infty (\bbZ^\nu)$ so that 
$\sigma (H_0+V)=[-2\nu, 2\nu]$ and so that $\limsup_{n\to\infty} \abs{V(n)} >0$. 
\end{theorem} 

We will also provide a new proof of Theorem~2. 

The key to the dimension dependence is the issue of finding $\varphi_n\in\ell^2 (\bbZ^\nu)$ so that
$\varphi_n (0)=1$ and $\langle \varphi_n, (2\nu -H_0)\varphi_n\rangle \to 0$. We will see that this 
can be done in dimension $1$ and $2$. It cannot be done in three or more dimensions, essentially because 
$(2\nu -H_0)^{-1}$ exists, not as a bounded operator on $\ell^2$ but as a matrix defined on vectors of 
finite support. To minimize $\langle\varphi, (2\nu -H_0)\varphi\rangle$ subject to $\varphi (0)=1$, 
by the method of Lagrange multipliers, one takes $\wti\varphi =(2\nu-H_0)^{-1}\delta_0/\langle\delta_0 
(2\nu -H_0)^{-1} \delta_0\rangle$. This is not in $\ell^2$ but has $\ell^2$ approximations. In fact, 
let $\varphi\in\ell^2$ with $\varphi(0)=\langle\delta_0, \varphi\rangle =1$. By the Cauchy-Schwarz 
inequality, $1\leq \|(2\nu - H_0)^{1/2} \varphi\|\,\|(2\nu - H_0)^{-1/2} \delta_0\|$, that is, 
\[
\langle \varphi, (2\nu-H_0)\varphi\rangle \geq \langle \delta_0, (2\nu - H_0)^{-1}\delta_0\rangle^{-1} 
>0
\]
for $\nu\geq 3$. So any $\ell^2$ sequence $\varphi$ with $\varphi(0)=1$ has a minimal kinetic energy in 
dimension $\nu\geq 3$. 

A different way of thinking about this is as follows: If $\varphi$ has compact support in a box of size 
$L$ and $\varphi (0)=1$, then, on average, $\nabla\varphi$ is at least $L^{-1}$ so $\|\nabla\varphi\|^2 
= \langle\varphi, (2\nu -H_0)\varphi\rangle\sim L^\nu L^{-2}$. If $\nu\geq 3$, one does not do better 
by taking big boxes. In $\nu=1$, one certainly does; and in $\nu=2$, a careful analysis will give 
$(\ln L)^{-1}$ decay. 

\begin{proposition}\lb{P4.3} Let $L_1, L_2 \geq 1$. There exists $\varphi_{L_1, L_2}\in\ell^2(\bbZ)$,
supported in $[-L_1, L_2]$, so that 
\begin{SL} 
\item[{\rm{(i)}}] $\varphi_{L_1, L_2}(0)=1$
\item[{\rm{(ii)}}] $\langle\varphi_{L_1, L_2}, (2-H_0) \varphi_{L_1, L_2}\rangle = (L_1 + 1)^{-1} + (L_2 +1)^{-1}$
\item[{\rm{(iii)}}] for suitable constants $c_1>0$ and $c_2 <\infty$,
\begin{equation} \lb{4.2}
c_1 (L_1 + L_2) \leq \|\varphi_{L_1, L_2}\|^2 \leq c_2 (L_1 + L_2)
\end{equation}
\end{SL}
\end{proposition} 

\begin{proof} Define
\begin{equation} \lb{4.3}
\varphi_{L_1, L_2}(n) = \begin{cases} 
1-\f{n}{L_2+1} & 0\leq n \leq L_2 +1 \\
1 - \f{\abs{n}}{L_1 +1} & 0\leq -n\leq L_1 +1 \\
0 & n\geq L_2 +1 \text{ or } n\leq -L_1 -1 \end{cases}
\end{equation}
then (i) and (iii) are easy.  As
\begin{equation} \lb{4.4}
\langle\psi, (2-H_0)\psi\rangle =
\sum_{j=-\infty}^\infty \big[\psi(j+1) - \psi(j)\big]^2
\end{equation}
for any $\psi\in \ell^2 (\bbZ)$, we have 
\begin{align*} 
\langle\varphi_{L_1, L_2}, (2-H_0)\varphi_{L_1, L_2}\rangle &=\sum_{j=1}^{L_2+1} 
\bigl( \tfrac{1}{L_2 +1}\bigr)^2 + \sum_{j=-1}^{L_1 +1} \bigl( \tfrac{1}{L_1 +1}\bigr)^2 \\
&= (L_1 +1)^{-1} + (L_2 +1)^{-1}
\end{align*} 
which proves (ii).
\end{proof} 

{\it Remark.} If $\psi(0)=1$ and $\psi$ is supported in $[-L_1, L_2]$,
$$
\sum_{j=1}^{L_2 +1} \psi (j) - \psi (j-1)= -1
$$
so, by the Schwarz inequality, 
\[
1 \leq (L_2 +1) \sum_{j=1}^{L_2 +1} \, \abs{\psi(j) - \psi(j-1)}^2 
\]
Thus 
\[
\langle\psi, (2-H_0)\psi\rangle \geq (L_1 +1)^{-1} + (L_2 +1)^{-1}
\]
which shows that \eqref{4.3} is an extremal function. 

\begin{proposition}\lb{P4.4} Let $L\geq 1$. There exists $\varphi_L\in\ell^2 (\bbZ^2)$ supported in 
$\{(n_1, n_2)\mid \abs{n_1}+\abs{n_2}\leq L\}$ so that 
\begin{SL}
\item[{\rm{(i)}}] $\varphi_L(0)=1$ 
\item[{\rm{(ii)}}] $0\leq \langle\varphi_L, (4-H_0)\varphi_L\rangle \leq c [\ln (L+1)]^{-1}$ for some $c>0$ 
\item[{\rm{(iii)}}] $(L^{-1} \ln(L))^2 \|\varphi_L\|^2 \to d >0$ 
\end{SL} 
\end{proposition} 

{\it Remark.} It seems clear that one cannot do better than $\ln (L)^{-1}$ in the large $L$ asymptotics 
of $\langle\varphi_L, (4-H_0)\varphi_L\rangle$ for any test function obeying (i) and the support 
condition. 

\begin{proof} Define 
\[
\varphi_L(n_1, n_2) = \begin{cases} 
\f{-\ln [(1+\abs{n_1} + \abs{n_2})/(L+1)]}{\ln(L+1)} & \text{if } \abs{n_1} + \abs{n_2} \leq L \\
0 & \text{if } \abs{n_1} + \abs{n_2} \geq L 
\end{cases} 
\]
then (i) is obvious. As
\begin{align*}
\ln\biggl(\f{a+1}{(L+1)}\biggr) - \ln \biggl( \f{a}{(L+1)}\biggr) &= \ln \biggl( 1+\f{1}{a}\biggr) \leq a^{-1} 
\end{align*}
we have that 
\begin{align*} 
\langle\varphi_L, (4-H_0)\varphi_L\rangle &= \sum_{n_1, n_2} \bigl(\varphi_L(n_1 +1, n_2) - \varphi_L (n_1, n_2)\bigr)^2 \\
&\qquad + \bigl(\varphi_L (n_1, n_2 +1) - \varphi_L (n_1, n_2)\bigr)^2 \\
&\leq \ln (L+1)^{-2} \sum_{\substack{ n_1, n_2 \\ \abs{n_1} + \abs{n_2} \leq L }} \, 
(1+ \abs{n_1} + \abs{n_2})^{-2} \\
&\leq c \ln (L+1)^{-1} 
\end{align*}
since the sum diverges as $\ln L$. This proves (ii). 

To prove (iii), we note that, by a simple approximation argument,  
\[
\ln(L)^2 L^{-2} \|\varphi_L\|^2 \to \iint_{\abs{x} + \abs{y} \leq 1} [\ln(\abs{x} + \abs{y})]^2 \, dx\,dy 
\]
as $L\to\infty$.
\end{proof}

\begin{proof}[Proof of Theorem~\ref{T4.1}] Consider first the case $\nu=1$. Suppose $\limsup \abs{V(n)} 
=a>0$. Pick $L$ so that $2(L+1)^{-1} < \f18 \min (a^2, 2a)$. Pick a sequence $n_1, \dots, n_j, \dots$ 
with $\abs{V(n_j)}\to a$ so that $\abs{n_j} -\max_{1\leq \ell \leq j-1} \abs{n_\ell} \geq 2(L+2)$. 
Thus, $\abs{n_j-n_\ell} \geq 2(L+2)$ for all $j\neq \ell$. 

Define 
\begin{equation} \lb{4.5}
F(n) = \min\biggl( 1, \f{2}{\abs{V(n)}}\biggr)
\end{equation}
and let $\psi_j(n)=\varphi_{L,L}(n-n_j)$. Then 
\begin{align*} 
\langle \psi_j, (H_0 -2+\tfrac14\, FV^2)\psi_j\rangle &\geq -2(L+1)^{-1} + \tfrac14\, F(n_j) V(n_j)^2 \\
&\geq -\tfrac18\, \min (a^2, 2a) + \tfrac14\, \min (\abs{V(n_j)}^2, 2\abs{V(n_j)})
\end{align*}
Thus we have that 
\[
\liminf \langle \psi_j, (H_0 - 2+\tfrac14\, FV^2)\psi_j\rangle \geq \tfrac{1}{8}\,\min (a^2, 2a) 
\]

As $\abs{FV} \le 2$, if $\varphi_{\pm, j} = (1\pm \f14\, FV)\psi_j$, we have 
\begin{equation} \lb{4.6}
\tfrac12\, \|\psi_j\| \leq \|\varphi_{\pm, j}\| \leq \tfrac32\, \|\psi_j\| \leq C_L
\end{equation}
where $C_L$ is independent of $j$; compare \eqref{4.2}. 

By \eqref{2.9}, we have a subsequence of $j$'s so that either 
\[
\liminf \langle \varphi_{+,j_\ell}, (H_0 +V-2)\varphi_{+, j_\ell}\rangle \geq \tfrac1{16}\, \min(a^2, 2a) 
\]
or 
\[
\liminf \langle \varphi_{-,j_\ell}, (-H_0 -V-2)\varphi_{+,j_\ell}\rangle \geq \tfrac1{16}\, \min(a^2, 2a) 
\]
Moreover, the $\varphi$'s are orthogonal. Thus $H$ has essential spectrum in either 
\[
[2+\tfrac{1}{16}\, d^{-1} \min(a^2, 2a),\infty) \text{ or } (-\infty, -2 -\tfrac{1}{16}\, d^{-1} \min 
(a^2, 2a)]
\]

The proof for $\nu=2$ is similar, using Proposition~\ref{P4.4} in place of Proposition~\ref{P4.3}. 
\end{proof}

\begin{proof}[Proof of Theorem~\ref{T4.2}] We will give an example with $V\geq0$.
Thus the only spectrum that $H_0 +V$ can have outside $[-2\nu, 2\nu]$ is in $(2\nu, \infty)$.

As $\nu\geq 3$, the operator $(2\nu -H_0)^{-1}$ has finite matrix elements despite being unbounded.
We denote the $n,m$ matrix element, the Green function, by $G_\nu(n-m)$. 
By the Birman-Schwinger principle \cite[Section 3.5]{Thir}, if the matrix 
\[
M_{nm}= V(n)^{1/2} G_\nu (n-m) V(m)^{1/2}
\]
defines an operator on $\ell^2 (\bbZ^\nu)$ with norm strictly less than $1$, then $H_0+V$ has no  
spectrum in $(2\nu, \infty)$. 

Since $G_\nu (n)\to 0$ as $n\to\infty$ (indeed, it decays as $\abs{n}^{-(\nu-2)}$), we can find a 
sequence in $\bbZ^\nu$ with $\abs{n_j}\to\infty$ and 
\begin{equation} \lb{4.7}
\sum_{j\neq k} \;\abs{ G_\nu (n_j - n_k) } < \tfrac12
\end{equation}
For example, pick $n_k$ inductively so $\sum_{j<k} G_\nu (n_j - n_k) < 2^{-k-2}$.
(Actually, $G_\nu (n-m)>0$ for all $n$ and $m$ so the absolute value sign is redundant.)
Choose $\lambda >0$ so that 
\begin{equation} \lb{4.8}
\lambda G_\nu (0) < \tfrac12
\end{equation}
and define $V$ by 
\[
V(n) = \begin{cases} \min (1, \lambda) & n=\text{some } n_j \\
0 & \text{otherwise} \end{cases}
\]
In this way, $\limsup_{\abs{n}\to\infty} \abs{V(n)}=\min (1,\lambda) >0$.  However, by Schur's lemma,
$\| M\| <1$ so $H_0+V$ has no eigenvalues.
\end{proof} 

The ideas in the first part of this section allow us to reprove Theorem~2 and, more importantly, extend 
it to two dimensions. 

\begin{theorem} \lb{T4.5} Let $\nu=1$ or $2$. If $\sigma (H_0+V)\subset [-2\nu, 2\nu]$, then $V=0$. 
\end{theorem} 

\begin{proof} By Theorems~\ref{T4.1} and \ref{T4.2}, $V(n)\to 0$. By Theorem~\ref{T3.1}, if $H_0 +V$ 
has no bound states, neither does $H_0 + \f1{4 \nu} V^2$. Since $V=0$ if and only if $V^2 =0$, we 
may as well consider the case $V\geq 0$. Let $\varphi_L$ be the function guaranteed by 
Proposition~\ref{P4.3} or \ref{P4.4}. Then 
\[
\langle \varphi_L, (H_0 + V-2\nu)\varphi_L\rangle \geq V(0) + \langle \varphi_L, (H_0 -2\nu) 
\varphi_L\rangle 
\]
Since $\langle \varphi_L, (H_0-2\nu)\varphi_L\rangle \to 0$, we must have $V(0)=0$. By translation 
invariance, $V(n)=0$ for all $n$. 
\end{proof} 

\begin{theorem} \lb{T4.6} Let $J$ be the Jacobi matrix \eqref{1.3a}. Suppose $\liminf a_n \geq 1$ and 
$\sigma_{\ess} (J)\subset [-2,2]$. Then $b_n \to 0$ as $n\to\infty$. 
\end{theorem} 

\begin{proof} Since $\liminf a_n \geq 1$, we can suppose $a_n \geq 1$ since the change from $a_n$ to 
$\min (a_n, 1)$ is a compact perturbation. By the lemma below, $\sigma_{\ess}(J)$ can only shrink if 
$a_n \geq 1$ is replaced by $a_n=1$. Thus we can suppose $a_n=1$ in what follows. 

Let $\wti H=H_0$ on $\ell^2 (\bbZ\backslash\bbZ^+)$ with a Dirichlet boundary condition at $0$,
$\wti H=J$ on $\ell^2 (\bbZ^+)$, and
$$
V(n)=\begin{cases} 0 & n \leq 0\\ b_n & n\geq 1\end{cases}
$$ 
Then $H=H_0 +V$ differs from $\wti H$ by a finite rank perturbation. Thus $H$ has essential spectrum 
in $[-2,2]$. The proof is completed by using Theorem~\ref{T4.1}. 
\end{proof} 

\begin{lemma} \lb{L4.7} If $J(\{a_n\}, \{b_n\})$ is the Jacobi matrix given by \eqref{1.3a}, then $\sup 
\sigma_{\ess} (J(\{a_n\}, \{b_n\}))$ and $-\inf \sigma_{\ess} (J(\{a_n\}, \{b_n\}))$ are monotone 
increasing as $a_n$ increases. 
\end{lemma} 

\begin{proof} As noted in Section~3 of Hundertmark-Simon \cite{HS}, for each $N$, the sum of the $N$ largest positive eigenvalues, 
$\sum_{j=1}^N E_j^+ (J(\{a_n\}, 
\{b_n\}))$, is monotone in $\{a_n\}$. But 
\[
\sup \sigma_{\ess} \big(J(\{a_n\}, \{b_n\})\big)
= \lim_{n\to\infty} \f{1}{N} \sum_{j=1}^N E_j^+ \big(J(\{a_n\}, \{b_n\})\big)
\]
The proof for $-\inf \sigma_{\ess}$ is similar. 
\end{proof}

\bigskip

\section{Decay and Bound States for\\ 
Half-Line Discrete Schr\"odinger Operators} \lb{s5}

While whole-line discrete Schr\"odinger operators have bound states if $V\not\equiv 0$ (Theorem~2), 
this is not true for half-line operators. Indeed, the discrete analogue of Bargmann's bound \cite{HS} 
implies that 
\begin{equation} \lb{5.1}
\sum_{n=1}^\infty n\abs{V(n)} <1 \Rightarrow \sigma (J_0 + V) =[-2,2]
\end{equation}
where $J_0$ is the free Jacobi operator, that is, \eqref{1.3a} with $a_n \equiv 1$, $b_n \equiv 0$. 

One can also include the endpoint case: If a sequence of selfadjoint operators $A_k$ converges strongly to $A$, then
$$
\sigma(A) \subseteq \bigcap_n \;\overline{ \bigcup_{k\geq n} \sigma(A_k) }
$$
see \cite[Theorem VIII.24]{rs1}.  This shows that (\ref{5.1}) can be extended to
\begin{equation}\label{5.1a}
\sum_{n=1}^\infty n\abs{V(n)} \leq 1 \Rightarrow \sigma (J_0 + V) =[-2,2]
\end{equation}

\medskip

In this section, we explore what the absence of bound states tells us about the decay of $V$\!. We begin 
with the case $V\geq 0$: 

\begin{theorem}\lb{T5.1} Suppose $V(n)\geq 0$ and that $J_0 +V$ has no bound states. Then 
\begin{equation} \lb{5.2}
\abs{V(n)}\leq n^{-1}
\end{equation}
Moreover, \eqref{5.2} cannot be improved in that for each $n_0$, there exists $V_{n_0}$ so that $V_{n_0} 
(n_0)=n_0^{-1}$ and $J_0 +V_{n_0}$ has no bound states. 
\end{theorem} 

\begin{proof} Let $W_{n_0}$ be 
\[
W_{n_0}(n) = \begin{cases} 1 & n=n_0 \\
0 & n\neq n_0 \end{cases} 
\]

We claim $J_0 + \lambda W_{n_0}$ has a bound state if and only if $\abs{\lambda} >n_0^{-1}$. By 
\eqref{1.7}, we can suppose $\lambda >0$. In that case, by a Sturm oscillation theorem \cite{Tes}, 
there is a bound state in $(2,\infty)$ if and only if the solution of 
\begin{equation} \lb{5.3}
u(n+1) + u(n-1) + \lambda W_{n_0} (n) u(n) = 2u(n) \qquad u(0)=0, \, u(1)=1
\end{equation}
has a negative value for some $n\in\bbZ^+$. The solution of \eqref{5.3} is 
\[
u(n) =\begin{cases} n & n\leq n_0 \\ 
n_0 + (1-\lambda n_0) (n-n_0) & n\geq n_0 \end{cases} 
\]
which takes negative values if and only if $\lambda n_0 >1$. This proves the claim. 

In particular, $n_0^{-1} W_{n_0} = V_{n_0}$ is a potential where equality holds in \eqref{5.2} and 
$\sigma (J_0+V_0) =[-2,2]$. 

On the other hand, if $V(n_0) > n_0^{-1}$, then since $V\geq 0$, $V(n) \geq V(n_0) W_{n_0}(n)$ for all 
$n$ and so, by a comparison theorem and the fact that we have shown $J_0 + V(n_0) W_{n_0}$ has a bound 
state, we have that $J_0 +V$ has a bound state. The contrapositive of $V(n_0)>n_0^{-1}\Rightarrow \sigma 
(J_0 +V)\neq [-2,2]$ is the first assertion of the theorem. 
\end{proof} 

{\it Remark.} Notice that Theorem~\ref{T5.1} says \eqref{5.1a} is optimal in the very strong sense that 
if $\sum_{n=1}^\infty \alpha_n \abs{V(n)} \leq 1\Rightarrow \sigma (J_0+V)=[-2,2]$ for all potentials $V$\!, 
then each $\alpha_n \leq n$.

\smallskip
Positivity of the potential made the proof of Theorem~\ref{T5.1} elementary. Because of the magic of Theorem~5, we can deduce a 
result for $V$'s of arbitrary sign:

\begin{theorem}\lb{T5.2} If $J_0 +V$ has no bound states, then 
\begin{equation} \lb{5.4}
\abs{V(n)} \leq 2n^{-1/2}
\end{equation}
Moreover, \eqref{5.4} cannot be improved by more than a factor of $2$ in that for each $n_0$, there 
exists $V_{n_0}$ so that $J_0 + V_{n_0}$ has no bound states and 
\[
\lim_{n_0\to\infty} \, n_0^{1/2} \abs{V_{n_0}(n_0)}=1
\]
\end{theorem} 

{\it Remarks.} (a) The proof shows 
\[
V_{n_0} (n_0) = \sqrt{\tfrac{1}{n_0} + \tfrac{1}{4n_0^2}\,} - \tfrac{1}{2n_0} \equiv \beta_{n_0}
\]
so \eqref{5.4} cannot be improved to value better than $\beta_{n_0} \sim n_0^{-1/2} - \tfrac12 n_0^{-1}$.

\smallskip
(b) In \cite{DK} it is shown that the absence of bound states implies
$$
\abs{V(n)} \le \sqrt{2} n^{-1/2}(1 + \tfrac2n)^{3/2}
$$
and that there are examples $V_{n_0}$ with $V_{n_0}(n_0) = \sqrt{2} n_0^{-1/2}$ and no bound states.

\begin{proof} Theorem~5 extends to the situation where $H_0$ is replaced by $J_0$ since the mapping 
$\varphi \to \varphi(1\pm FV)$ is local. Thus if $J_0 +V$ has no bound states, neither does $J_0 + 
\f14 V^2$. Since $V^2\geq 0$, Theorem~\ref{T5.1} applies, and thus $\f14 \abs{V(n)}^2 \leq n^{-1}$, 
which is \eqref{5.4}. 

For the other direction, let $W_{n_0}$ be 
\[
W_{n_0} = \begin{cases} 1 & n=n_0 \\
-1 & n=n_0 +1 \\ 
0 & n\neq n_0, \, n_0 +1 \end{cases}
\]
A direct solution of \eqref{5.3} is 
\begin{equation} \lb{5.5}
u(n) = \begin{cases} n & n\leq n_0 \\
(1 - \lambda) n_0 + 1 + (1+\lambda -\lambda^2 n_0)(n-n_0-1)  & n\geq n_0 +1 \\
\end{cases}
\end{equation}

Thus $u(n)$ has a negative value if and only if $1+\lambda -\lambda^2 n_0 <0$. Define 
\begin{equation} \lb{5.6}
\lambda_\pm^{\crit} = \pm \sqrt{\tfrac{1}{4n_0^2} + \tfrac{1}{n_0}\,} - \tfrac{1}{2n_0}
\end{equation}
If $\abs{\lambda} >\min (\abs{\lambda_+^{\crit}},\abs{\lambda_-^{\crit}})$, $u$ takes negative values 
for either $u(n,\lambda)$ or $u(n, -\lambda)$. By \eqref{1.7}, $J_0 +V$ has eigenvalues in $(-\infty, 
-2)$ if and only if $J_0 -V$ has eigenvalues in $(2,\infty)$. Thus since $\abs{\lambda_+^{\crit}} < 
\abs{\lambda_-^{\crit}}$, $J_0 + \lambda W_{n_0}$ has no eigenvalues if $\abs{\lambda} \leq 
\lambda_+^{\crit}$. 
\end{proof} 

One can also say something about infinitely many bound states:

\begin{theorem}\lb{T5.3} 
\begin{SL} 
\item[{\rm{(i)}}] If $V \geq 0$ and 
\begin{equation} \lb{5.7}
\limsup_{n\to\infty} \, \abs{V(n)}n >1
\end{equation}
then $J_0 +V$ has infinitely many bound states. 
\item[{\rm{(ii)}}] For general $V$\!, if $\limsup_{n\to\infty}\abs{V(n)}n^{1/2} >2$, then $J_0 +V$ 
has infinitely many bound states. 
\end{SL} 
\end{theorem} 

\begin{proof} (ii) follows from (i) by Theorem~5. To prove (i), suppose $J_0 +V$ has only finitely 
many bound states. Then $(J_0 +V-2)u$ has only finitely many sign changes, so there is $N_0$ with 
$u(n) u(n+1)>0$ if $n>N_0$. It follows that $J_0 +V$ with $\wti V(n) = V(n+N_0)$ has no bound states. 
Thus $\abs{\wti V(n)}\leq n^{-1}$, so $\limsup_{n\to\infty} n\abs{V(n)} \leq 1$. Thus, by contrapositives, 
\eqref{5.7} implies $J_0 + V$ has infinitely many bound states. 
\end{proof} 

\begin{example}\lb{E5.4}
Let $N$ be a positive integer and $n_k = N^{2k}$. We consider the sequence $u(n)$ which has slope $u(n+1) - u(n) = N^{-k}$ for  $n \in 
[n_k,n_{k+1})$ and then determine the potential $V$ at the sites $n_k$ so that $u$ is the generalized eigenfunction at energy $2$. (Constancy of the 
slope in the 
intervals $(n_k,n_{k+1})$ implies that the potential vanishes there.) We have
\begin{align*}
u(n_k) &= n_1 + (n_2 - n_1) N^{-1} + \cdots + (n_k - n_{k-1}) N^{-(k-1)}\\
&= (1 - N^{-1}) \{ N^2 + N^3 + \cdots + N^k \} + N^{k+1}\\
&= N^{k+1} \{ 1 + N^{-1} - N^{-k} \}
\end{align*}
and so
\begin{align*}
V(n_k)& = \frac{2 u(n_k) - u(n_k + 1) - u(n_k - 1)}{u(n_k)}\\
& = \frac{N^{1-k} - N^{-k}}{N^{k+1} \{ 1 + N^{-1} - N^{-k} \}}\\
& = \frac{1 - N^{-1}}{1 + N^{-1} - N^{-k}} \, \frac{1}{n_k}
\end{align*}

As $u$ is monotone, there are no sign flips. We may conclude that $J_0+V$ has no bound states because $V(n) \ge 0$.
Therefore, taking $N \to \infty$, we see that the $1$ in \eqref{5.7} is optimal.

A similar argument \cite{Zl} shows there are examples with $\limsup n^{1/2} \abs{V(n)}=1-\veps$ and 
no bound states for each $\veps > 0$. Basically, $V(n)\neq 0$ for $n=n_k$ or $n_k +1$ and $V(n_k) =-V(n_k +1)=n_k^{-1/2} 
(1-\veps_k)$ with $\veps_k \to \veps$. Again, $n_k$ must grow at least geometrically. \qed
\end{example}

The examples that saturate Theorems~\ref{T5.1} and \ref{T5.3} are sparse, that is, mainly zero. If $V$ 
is mainly nonzero and comparable in size, the borderlines change from $n^{-1}$ to $n^{-2}$ for positive 
$V$'s and from $n^{-1/2}$ to $n^{-1}$ for $V$'s of arbitrary sign. 

\begin{theorem}\lb{T5.5} Let $V\geq 0$. Suppose there exists $\veps >0$ and $n_k\to\infty$ so that 
\begin{SL} 
\item[{\rm{(i)}}] 
\begin{equation} \lb{5.8}
\f{2}{n_k} \sum_{j=n_k/2}^{n_k} \, V(j) \geq \veps V(n_k)
\end{equation}
\item[{\rm{(ii)}}] $\limsup_{k\to\infty} \veps n_k^2 V(n_k) > 48$ 
\end{SL}
Then $J_0+V$ has infinitely many bound states. 
\end{theorem} 

\begin{proof} For notational simplicity, we suppose each $n_k$ is a multiple of $4$. By passing to a 
subsequence, we can suppose that 
\begin{gather} 
\f{\veps n_k}{8}\, V(n_k) > \f{6}{n_k} \lb{5.9} \\ 
\f{n_{k+1}}{4} > \f{3}{2}\, n_k + 2 \lb{5.10}
\end{gather} 

Let $u_k$ be the function which is $1$ at $n_k$, has constant slope on the intervals $[\f{n_k}{4}-1, n_k]$ and 
$[n_k, \f{3n_k}{2} + 1]$, and vanishes at $n=\f{n_k}{4}-1$ and $n=\f{3n_k}{2}+1$. By Proposition~\ref{P4.3}, 
\[
\langle u_k, (2-J_0)u_k\rangle \leq \f{6}{n_k}
\]

On $[\f{n_k}{2}, n_k]$, we have $\abs{u(j)}\geq \f12$, so 
\begin{align*} 
\langle u_k, Vu_k\rangle &\geq \tfrac14\sum_{j=n/2}^n \, V(j) \geq \f{\veps n_k}{8}\, V(n_k) \qquad \text{(by \eqref{5.8})} 
\end{align*}

By \eqref{5.9}, $\langle u_k, (J_0 + V-2)u_k\rangle >0$ for all $k$. By \eqref{5.10} for $k\neq \ell$, 
\[
\langle u_k, u_\ell\rangle = \langle u_k, (J_0 +V)u_\ell\rangle =0
\]
so, by the min-max principle, $J_0 +V$ has infinitely many eigenvalues in $(2,\infty)$. 
\end{proof}

Theorem~5 and Theorem~\ref{T5.5} immediately imply

\begin{theorem}\lb{T5.6} Suppose there exists $\veps >0$ and $n_k\to\infty$ so that 
\begin{SL} 
\item[{\rm{(i)}}] $\f{2}{n_k} \sum_{j=n_k/2}^{n_k} \abs{V(j)}^2 \geq \veps^2 \abs{V(n_k)}^2$ 
\item[{\rm{(ii)}}] $\limsup_{k\to\infty} \veps n_k \abs{V(n_k)} > 8\sqrt{3}$ 
\end{SL}
Then $J_0 +V$ has infinitely many bound states. 
\end{theorem} 

In this regard, here is another application of Theorem~5: 

\begin{theorem}\lb{T5.7} If $\abs{V(n)} \ge \f{\beta}{n}$ with $\beta > 1$ and $V(n) \to 0$, then $J_0+V$ has infinitely 
many bound states.
\end{theorem} 

\begin{proof} It is known (see \cite[Theorem A.7]{DHS}) if $\beta^2 >1$, then the operator with potential 
$\f{\beta^2}{4n^2}$, and hence the operator with potential $\f14 V(n)^2 \ge \f{\beta^2}{4n^2}$, has infinitely many bound states.
The assertion now follows from Theorem~5.
\end{proof} 

\begin{corollary}
If $V(n) \to 0$ but $\liminf_{|n| \to \infty} |n V(n)| > 1$, then $J_0+V$ has infinitely
many bound states. The same result holds in the whole-line setting.
\end{corollary}

\begin{proof}
We begin with the half-line case. By hypothesis, there exists a $\beta > 1$ such that $\abs{V(n)} \ge \f{\beta}{n}$ for all 
but finitely many $n$. Therefore the claim follows from the previous theorem because a finite rank perturbation can remove 
at most finitely many eigenvalues. The whole-line case follows by Dirichlet decoupling.
\end{proof}

{\it Remark.} It is known (see \cite{DHS}) that if $V(n) = \f{1}{4n^2}$ or $V(n) =
\beta \f{(-1)^n}{n}$ with $\abs{\beta}<\f12$, then $J_0 +V$ has finitely many bound states. Thus
the powers $n^{-2}$ and $n^{-1}$ in the previous results are optimal.

\medskip

The optimal constant in Theorem \ref{T5.7} is $1$, as we now show.

\begin{proposition}\label{P:AltEx}
For $\beta\in[-1,1]$, the operator $J_0+V$ with potential $V(n)=\beta \f{(-1)^n}{n}$ has no bound states.
\end{proposition}

\begin{proof}
We will show that the operator with potential $V(n)=\frac{(-1)^n}{n}$ has no bound states.  As the absolute value of a bound state
eigenvalue is an increasing function of the coupling constant, this implies that potentials of the form
$V(n)=\beta\f{(-1)^n}{n}$ have no bound states for $\beta\in[0,1]$.  Equation \eqref{2.5} shows that
$J_0+V$ is unitarily equivalent to $-(J_0-V)$.  Thus, the proposition for $\beta\in[-1,0]$ follows from
the $\beta\in[0,1]$ case. 

By the unitary equivalence of $J_0 + V$ and $-(J_0 - V)$, it suffices to show that 
for $V_0 = (-1)^n/n$, $J_0 + V_0$ and $J_0 - V_0$ have no eigenvalues in $(2,\infty)$. 

We look at solutions of 
\begin{equation} \lb{5.11a}
u(n+1) + u(n-1) = (2\mp V_0(n)) u(n) 
\end{equation}
By Sturm oscillation theory, the number of  
eigenvalues of $J_0 \pm V_0$ in $(2,\infty)$ is equal to the number of zeros, in $(0,\infty)$,
of the linear interpolation of the generalized eigenfunction---that is, the solution of \eqref{5.11a}
with $u(0)=0$.
Moreover, the Sturm separation theorem implies that if (\ref{5.11a}) has a solution with $u(n)> 0$ for
$n=0,1,2,...$, then the generalized eigenfunction must be positive for $n\geq1$ (and not $\ell^2$; see
remark below).

We are able to write down positive solutions explicitly, but rather than pull such a rabbit out of a hat, 
we provide some explanation. Motivated by calculations in Maple, we look for solutions with $u(n) =
u(n+1)$ for either all odd $n$ or all even $n$. This is equivalent to asking if  
\begin{equation} \lb{5.11b} 
\left( \begin{array}{rr} 
x & -1 \\ 1 & 0 \end{array} 
\right) 
\left( \begin{array}{rr} 
y & -1 \\ 1 & 0 \end{array} 
\right) 
= 
\left( \begin{array}{cr} 
xy-1 & -x \\ y & -1 \end{array} 
\right) 
\end{equation}
has $\binom{1}{1}$ as an eigenvector. If this is true for $y=E-V(n)$, $x=E-V(n+1)$ for all odd 
(resp.~even) $n$, then the Schr\"odinger equation has a solution with $u(n)=u(n-1)$ for all odd 
(resp.~even) $n$, and for such $n$, 
\begin{equation}\lb{5.11c}
u(n+2) = [E-V(n) -1] u(n)
\end{equation} 
The matrix in \eqref{5.11b} has $\binom{1}{1}$ as an eigenvector if and only if
\begin{equation} \lb{5.11d} 
xy = x+y 
\end{equation} 
If $x=2+a$, $y=2+b$, then \eqref{5.11a} becomes 
\begin{equation} \lb{5.11e} 
ab = -a-b 
\end{equation} 
This is solved by $b=\f{1}{m}$, $a=-\f{1}{m+1}$ with $y-1 = 1+\f{1}{m}$. Since $-V(n)$ appears in the 
transfer matrix for $V_0$, we take $m=2n+1$, $n=0,1,2,\dots$ and find a solution with 
\[
u(0) = u(1) =1 \quad u(2n) = u(2n+1), \quad u(2n+2) = \bigl( 1+\tfrac{1}{2n+1}\bigr) u(2n) 
\]
which is a positive solution with $u(n)\to\infty$ as $n^{1/2}$ as $n\to\infty$. For $-V_0$, we take 
$m=2n$, $n=1,2,\dots$, and find a solution with 
\[
u(0)=0 \quad u(1)=u(2)=1 \quad u(2n)=u(2n-1) \quad u(2n+2)=\bigl( 1 + \tfrac{1}{2n}\bigr) u(2n)
\]
so again, $u(n)\to\infty$ as $n^{1/2}$. We have thus found the required solution to show $J_0 +V_0$ 
has no eigenvalues in $(2,\infty)$.
\end{proof}

{\it Remarks.} (a) It follows from the proof that the generalized eigenfunctions at energies $\pm2$ are not
square summable.  This shows that $\pm2$ are not eigenvalues.

\smallskip

(b) Choosing $y=-\f{1}{m}$, $x=\f{1}{m+1}$ in the arguments given above shows that 
there are solutions $u_\pm$ of $(J_0 +V_0)u=0$ with $\abs{u_\pm(n)}\sim\abs{n}^{\pm 1/2}$ as $n\to\infty$.
This shows that $0$ is not an eigenvalue of $J_0 +V_0$ but suggesting that for $J_0 + (1+\veps)V_0$,
there are solutions $\ell^2$ at infinity for $\veps >0$.
That is, just as coupling $1$ is the borderline for eigenvalues outside $[-2,2]$, it is 
the borderline for an eigenvalue at $E=0$ similar to the Wigner-von Neumann phenomenon. 

\smallskip

As our final topic, we want to discuss divergence of eigenvalue moments if $|V(n)| \sim n^{-\alpha}$ 
with $\alpha <1$. 

\begin{lemma} \lb{L5.8} Let $A$ be a bounded selfadjoint operator. Let $\{\varphi_j\}_{j=1}^\infty$ 
be an orthonormal set with 
\begin{equation} \lb{5.11}
\langle \varphi_j, A\varphi_k\rangle = \alpha_j \delta_{jk}
\end{equation}
If $F$ is a nonnegative even function on $\bbR$ that is monotone nondecreasing on $[0,\infty)$, then
\begin{equation} \lb{5.12}
\tr \big(F(A)\big) \geq \sum_j F(\alpha_j)
\end{equation}
\end{lemma} 

{\it Remarks.} (a) As $F(A)\geq 0$, it follows that $\tr (F(A))$ is always defined although it may be infinite.

\smallskip
(b) In particular, if $\varphi_j$ is a family of nonzero vectors in $\ell^2(\bbZ^+)$ with
$\dist (\supp (\varphi_j),\supp(\varphi_k)) \geq 2$ for $j\neq k$, then for $J=J_0 + V$\!, 
\begin{equation} \lb{5.13}
\tr \big(F(J)\big) \geq \sum_j F \biggl(\biggl| \f{\langle \varphi_j, J\varphi_j \rangle}{\langle \varphi_j, 
\varphi_j\rangle} \biggr| \biggr)
\end{equation}

\begin{proof} Let $E_1 \geq E_2\geq \cdots$ be the eigenvalues of $\abs{A}$. By min-max and max-min 
for $A$, we have $E_j \geq \abs{\alpha_j}^*$ where $\abs{\alpha_j}^*$ is the decreasing rearrangement of 
$\abs{\alpha_j}$. So \eqref{5.12} follows. 
\end{proof} 

\begin{lemma} \lb{L5.9} Let $\abs{V}\leq 4\nu$ on $\supp (\varphi)$. Then there exists $\psi$ with 
$\supp (\psi)=\supp(\varphi)$ so that 
\begin{equation} \lb{5.14x}
\|\psi\|^{-2} \big\lvert \langle \psi, (H_0 +V)\psi\rangle \big\rvert - 2\nu \geq \tfrac14\, 
\bigl[\|\varphi\|^{-2} \langle \varphi, (H_0 + \tfrac1{4\nu}\, V^2)\varphi\rangle - 2\nu \bigr]
\end{equation}
\end{lemma} 

\begin{proof} Let $\psi_\pm = (1\pm (4\nu)^{-1}V)\varphi$. Since $\abs{V}\leq 4\nu$, $\|\psi_\pm\|^2 
\leq 4 \|\varphi\|^2$. The result now follows from \eqref{2.9} by choosing $\psi$ to be either $\psi_{+}$
or $U\psi_{-}$.
\end{proof} 

\begin{theorem} \lb{T5.10} Let $J$ be a Jacobi matrix of the form $J_0 +V$ where 
\begin{equation} \lb{5.14}
\abs{V(n)} \geq Cn^{-\alpha}
\end{equation}
for some $\alpha <1$ and $V(n) \to 0$. Then
\begin{equation} \lb{5.15}
\sum_j \big( |E_j| - 2 \big)^\gamma =\infty
\end{equation}
for 
\begin{equation} \lb{5.16}
\gamma < \f{1-\alpha}{2\alpha}
\end{equation}
where $E_j$ are eigenvalues of $J$ outside $[-2,2]$. 
\end{theorem} 

{\it Remark.} In particular, the eigenvalue sum $\sum_{j=1}^\infty (\abs{E_j}-2)^{1/2}$ critical of 
Szeg\H{o}-type sum rules \cite{KS,SZ} diverges if $\alpha <\f12$. This illuminates results in 
\cite{DHS,SZ}. 

\begin{proof} Fix $p>0$. Let $\varphi_m$ be supported near $m^{p+1}$ on an interval $[m^{p+1} - 
C_1 m^p, m^{p+1} + C_1 m^p]$ where $C_1$ is picked to arrange that supports are separated by at least 
$2$. Taking the slopes fixed on each half-interval and using Proposition~\ref{P4.3}, we see
\begin{align}
\langle \varphi_m, (2-H_0)\varphi_m \rangle &\leq \f{C_2}{m^p} \lb{5.17} \\
\langle \varphi_m, \tfrac14\, V^2 \varphi_m \rangle &\geq \f{C_3 m^p}{m^{2\alpha (p+1)}} \lb{5.18} \\
\langle \varphi_m, \varphi_m\rangle &\geq C_4 m^p \lb{5.19} 
\end{align}

So long as $\alpha (p+1)<p$ (i.e., $p<\f{\alpha}{1-\alpha}$), \eqref{5.18} beats out \eqref{5.17} for 
large $m$, and we find 
\begin{equation} \lb{5.19x}
\langle \varphi_m, \varphi_m \rangle^{-1} \langle \varphi_m, (H_0 + \tfrac14\, V^2 -2)\varphi_m 
\rangle \geq C_5 m^{-2\alpha (p+1)}
\end{equation}
As $p\downarrow \f{\alpha}{1-\alpha}$, $2\alpha (p+1) \downarrow \f{2\alpha}{1-\alpha}$. 

By the lemma with $F(x) =\dist (x, [-2,2])^\gamma$, we see that we have divergence if \eqref{5.16} holds. 
\end{proof} 

{\it Remarks.} (a) If the constant $C$ in \eqref{5.14} is large enough, we can take $p=\f{\alpha}{1-\alpha}$ 
and get divergence if $\gamma = \f{1-\alpha}{2\alpha}$. 

\smallskip
(b) One can extend this result as well as Theorems~\ref{T5.3} and \ref{T5.5} to higher dimensions.

\bigskip


\end{document}